\documentclass[reprint,superscriptaddress,aps,prl]{revtex4-1}
\usepackage{graphicx} 
\usepackage{notes2bib}

\newcommand{\uv}[1]{\ensuremath{\mathbf{\hat{#1}}}} 
 
\let\baraccent=\= 
\renewcommand{\=}[1]{\stackrel{#1}{=}} 

\graphicspath{{/Users/ryanday/Documents/UBC/Drafts/Spin_FeSCs/RPDay_FeSCs_PRL/v2_ref/PRL_V2_Figs/}}

\begin{document}

\title{Influence of Spin-Orbit Coupling in Iron-Based Superconductors}

\author{R.P. Day}
\author{G. Levy}
\affiliation{Department of Physics and Astronomy, University of British Columbia, Vancouver, BC V6T 1Z1, Canada}
\affiliation{Quantum Matter Institute, University of British Columbia, Vancouver, BC V6T 1Z4, Canada}
\author{ M. Michiardi}
\affiliation{Department of Physics and Astronomy, University of British Columbia, Vancouver, BC V6T 1Z1, Canada}
\affiliation{Quantum Matter Institute, University of British Columbia, Vancouver, BC V6T 1Z4, Canada}
\affiliation{Max Planck Institute for Chemical Physics of Solids, Dresden, Germany}
\author{B. Zwartsenberg}
\author{M. Zonno}
\author{F. Ji}
\author{E. Razzoli}
\author{F. Boschini}
\author{S. Chi}
\author{R. Liang}
\affiliation{Department of Physics and Astronomy, University of British Columbia, Vancouver, BC V6T 1Z1, Canada}
\affiliation{Quantum Matter Institute, University of British Columbia, Vancouver, BC V6T 1Z4, Canada}
\author{P.K. Das}
\affiliation{Istituto Officina dei Materiali (IOM)-CNR, Laboratorio TASC, Area Science Park, S.S.14, Km 163.5, I-34149 Trieste, Italy}
\affiliation{International Centre for Theoretical Physics (ICTP), Strada Costiera 11,
I-34100 Trieste, Italy}
\author{I. Vobornik}
\author{J. Fujii}
\affiliation{Istituto Officina dei Materiali (IOM)-CNR, Laboratorio TASC, Area Science Park, S.S.14, Km 163.5, I-34149 Trieste, Italy}
\author{W.N. Hardy}
\author{D.A. Bonn}
\author{I.S. Elfimov}
\author{A. Damascelli\thanks{damascelli@physics.ubc.ca}}
\email{damascelli@phas.ubc.ca}

\affiliation{Department of Physics and Astronomy, University of British Columbia, Vancouver, BC V6T 1Z1, Canada}
\affiliation{Quantum Matter Institute, University of British Columbia, Vancouver, BC V6T 1Z4, Canada}

\begin{abstract}
We report on the influence of spin-orbit coupling (SOC) in the Fe-based superconductors (FeSCs) via application of circularly-polarized spin and angle-resolved photoemission spectroscopy (CPS-ARPES). We combine this technique in representative members of both the Fe-pnictides (LiFeAs) and Fe-chalcogenides (FeSe) with tight-binding calculations to establish an ubiquitous modification of the electronic structure in these materials imbued by SOC. At low energy, the influence of SOC is found to be concentrated on the hole pockets, where the largest superconducting gaps are typically found. This effect varies substantively with the $k_z$ dispersion, and in FeSe we find SOC to be comparable to the energy scale of orbital order. These result contest descriptions of superconductivity in these materials in terms of pure spin-singlet eigenstates, raising questions regarding the possible pairing mechanisms and role of SOC therein.
 
\end{abstract}
\maketitle



The electronic structure of the iron-based superconductors (FeSCs) is characterized by several shallow Fermi-surface pockets which render the low-energy electronic structure susceptible to small interactions such as orbital order and spin-orbit coupling (SOC) \cite{Hirsch,Hirsch_pairing,ChubRev,vanr}. However, due to the relative success of non-relativistic methods in capturing much of the electronic structure and phenomenology of the FeSCs, SOC has been largely neglected in the discussion of these materials. Recent experimental observations contest this simplification as the breaking of spin-rotational invariance measured via inelastic neutron scattering (INS) \cite{MaFeSe,INS_122,Wasser_INS}, anisotropies in the superconducting gap parameter \cite{FeSe_3D,Shin_Octet}, and topologically non-trivial surface states \cite{FeSe_TSS,Wang_FeSeTe,TSS_FeSC,TSS_LiFeAs}, all suggest the importance of SOC in the physics of these materials. This has been corroborated by observation of energy splittings in angle-resolved photoemission spectroscopy (ARPES) measurements on a variety of FeSCs consistent with SOC \cite{WKu,Borisenko,Johnson,Liu,Fanfarillo,Suzuki}. Interpretation of these splittings is however complicated by significant band and orbital-dependent renormalizations in ARPES on FeSCs, as well as the remarkably similar influence of nematic or orbital order on the dispersion near the Brillouin zone centre \cite{Fernandes}. 

\begin{figure}[b!]
\centering
\includegraphics[width=\linewidth]{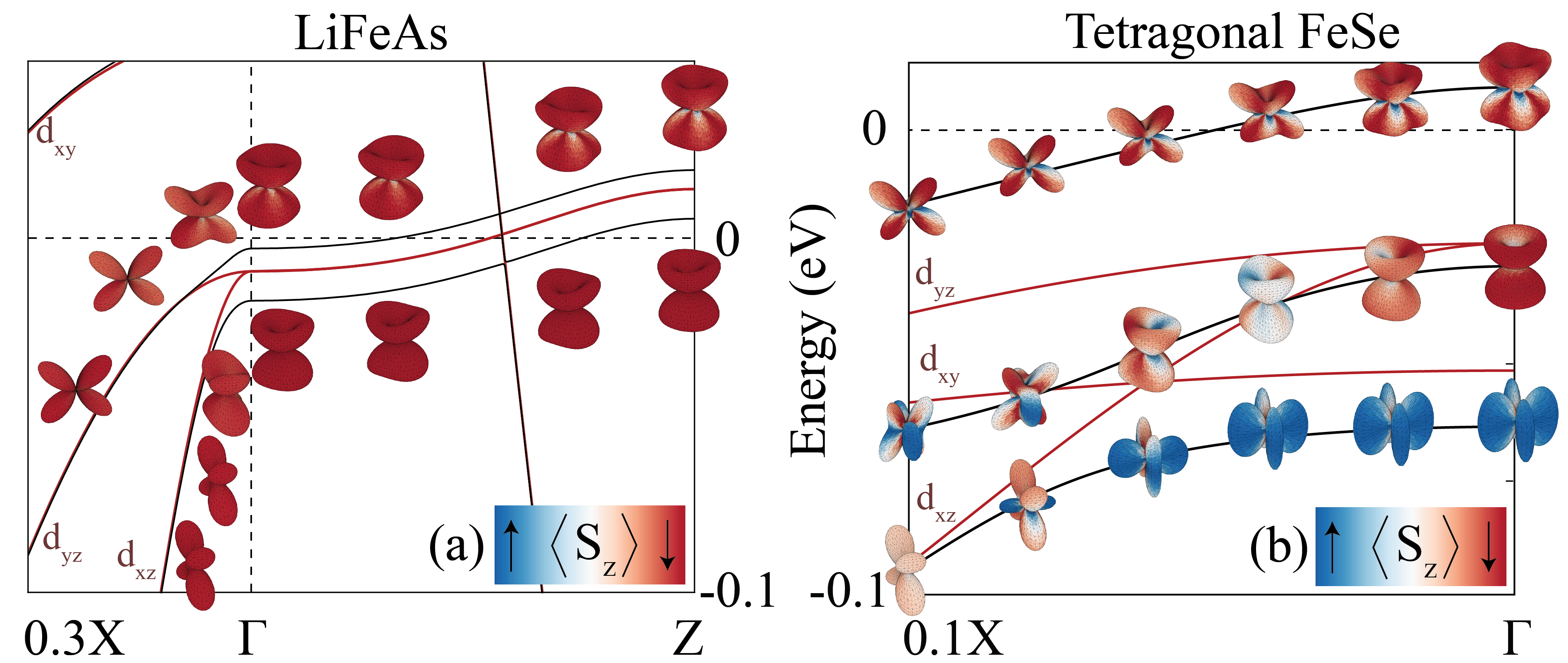}
\caption{Electronic structure with (black) and without (red) SOC for (a) LiFeAs and (b) FeSe. Orbitally projected eigenstates illustrate the substantial departure from cubic harmonics near the zone centre in both materials. $k_z$ dispersion for FeSe, not shown, is not markedly different from $\Gamma$. Colourscale indicates possible values of $\left<S_z\right>$. The red curves are labelled by their primary character--the $d_{xy}$ states play a critical role, as seen for the upper state in LiFeAs, and for both $d_{xz}$ and $d_{yz}$ states in FeSe, where the states can no longer be factorized into orbital and spin sectors. We note that each state is two-fold Kramers degenerate with a net spin of zero: the degenerate state of opposite spin is not shown for clarity.}
\end{figure}

	To provide a more comprehensive perspective on how SOC modifies the electronic structure of FeSCs throughout the Brillouin zone, we report here on the application of circularly-polarized spin (CPS)-ARPES to archetypal compounds LiFeAs and FeSe, exploring the entanglement of spin and orbital degrees of freedom for both in plane and perpendicular momentum. CPS-ARPES is an ideal probe for SOC, combining orbital-selectivity of circularly polarized light with spin detection to allow for direct and independent access to the spin and orbital vectors throughout the Brillouin zone, even in the absence of spin or charge order \cite{Veenstra,CRO,GaAs}. Our principal result is the observation of a strong entanglement of spin and orbital vectors out of plane in the vicinity of the Brillouin zone centre, which evolves towards the standard non-relativistic description only for larger in-plane momentum. This strong momentum-dependence reveals the relevance of the precise location of the chemical potential in determining the importance of SOC, as the spin-orbit entangled states can be pushed away from E$_F$ with doping. Furthermore, by studying FeSe in both the tetragonal and orthorhombic phase, we observe persistent entanglement of orbital and spin degrees of freedom in the presence of nematicity. 

	FeSe crystals were grown via vapour transport technique \cite{FeSe_growth} and LiFeAs by a self-flux method \cite{LiFeAs_growth}. Samples were cleaved and measured in the non-superconducting phase at 20 K at pressure of $10^{-10}$ mbar at the APE-LE endstation at ELETTRA using a Scienta DA30 analyzer (resolution set to 20 meV) equipped with a VLEED-based spin-detector (resolution set to 65 meV) \cite{APE}. The electronic structure calculations for LiFeAs are based on a 10 orbital tight-binding model adapted from Ref. \onlinecite{Eschrig} and \onlinecite{Hirsch_LiFeAs} to match the experimental spectra (see Fig. 2(a)) and detailed in the Supplementary Materials.
	
	Spin-orbit coupling leads to a significant departure of the electronic eigenstates near the Fermi level from the conventional description in terms of cubic-harmonics. At the Brillouin zone centre, rather than adhering to the conventional $d_{xz/yz}$ description, orbitals mix such that the orbital component is more readily described in terms of spherical harmonics $Y_{2}^{\pm1}$. In Fig. 1 we plot the orbitally-projected eigenstates for LiFeAs and FeSe at several points within the Brillouin zone near $E_F$, alongside possible spin-orientations; entangled relativistic orbitals dominate the low-energy electronic structure. Furthermore, the proximity of the $d_{xy}$ orbital introduces $L_xS_x$ and $L_yS_y$ terms, particularly affecting FeSe, as well as the upper state in LiFeAs. This is a direct consequence of SOC within the framework of a 2-Fe unit cell, as the 1-Fe unit cell has no $d_{xy}$ state in this region of energy and momentum space. By achieving an experimental measure of the entanglement of spin and orbital degrees of freedom in the FeSCs, we may establish a deeper understanding of the electronic states from which superconductivity and magnetism arise, and how the influence of SOC may carry the balance of power in establishing the low energy phase diagram in these materials. 
	
	Dipole selection rules associated with circularly polarized light of different helicity will photoemit preferentially from states of different $m_l$ projection. This allows the polarization helicity to act as an orbital filter on the photoemission process. A magnetized target in the VLEED detector then filters the photoelectrons according to their vectorial spin orientation \cite{APE}. Combining intensity maps with different polarization and spin-projections, we define the spin-polarization asymmetry \cite{Veenstra} as

\begin{equation}
P_i = \frac{\sqrt{I_-^{\uparrow}I_+^{\downarrow}}-\sqrt{I_+^{\uparrow}I_-^{\downarrow}}}{\sqrt{I_-^{\uparrow}I_+^{\downarrow}}+\sqrt{I_+^{\uparrow}I_-^{\downarrow}}}
\label{eq:Peq}
\end{equation}
where $I^{\uparrow(\downarrow)}_{+(-)}$ indicates the photocurrent intensity for $C_+$ ($C_-$) light incident on the $\uparrow$($\downarrow$) spin detector oriented along the $i=\uv{x},\uv{y},\uv{z}$ direction. The dipole selection rules above dictate that $I_-^{\uparrow}I_+^{\downarrow}$ is a measure of states with orbital and spin aligned parallel, and $I_+^{\uparrow}I_-^{\downarrow}$, those aligned antiparallel. Consequently, CPS-ARPES is the most direct measure of the effects of SOC, with $P_i$ offering an energy and momentum-resolved measure of the spin-orbit coupling polarization. In the absence of SOC, $C_{\pm}$ would photoemit from the orbitally-equivalent Kramers' degenerate spin states indiscriminately, resulting in a vanishing $P_i$. Measuring $P_i$ throughout the Brillouin zone and along different axes of spin projection, we may study the impact of spin-orbit coupling throughout the electronic structure of the FeSCs. 
\begin{figure}[t!]
\includegraphics[width=\linewidth]{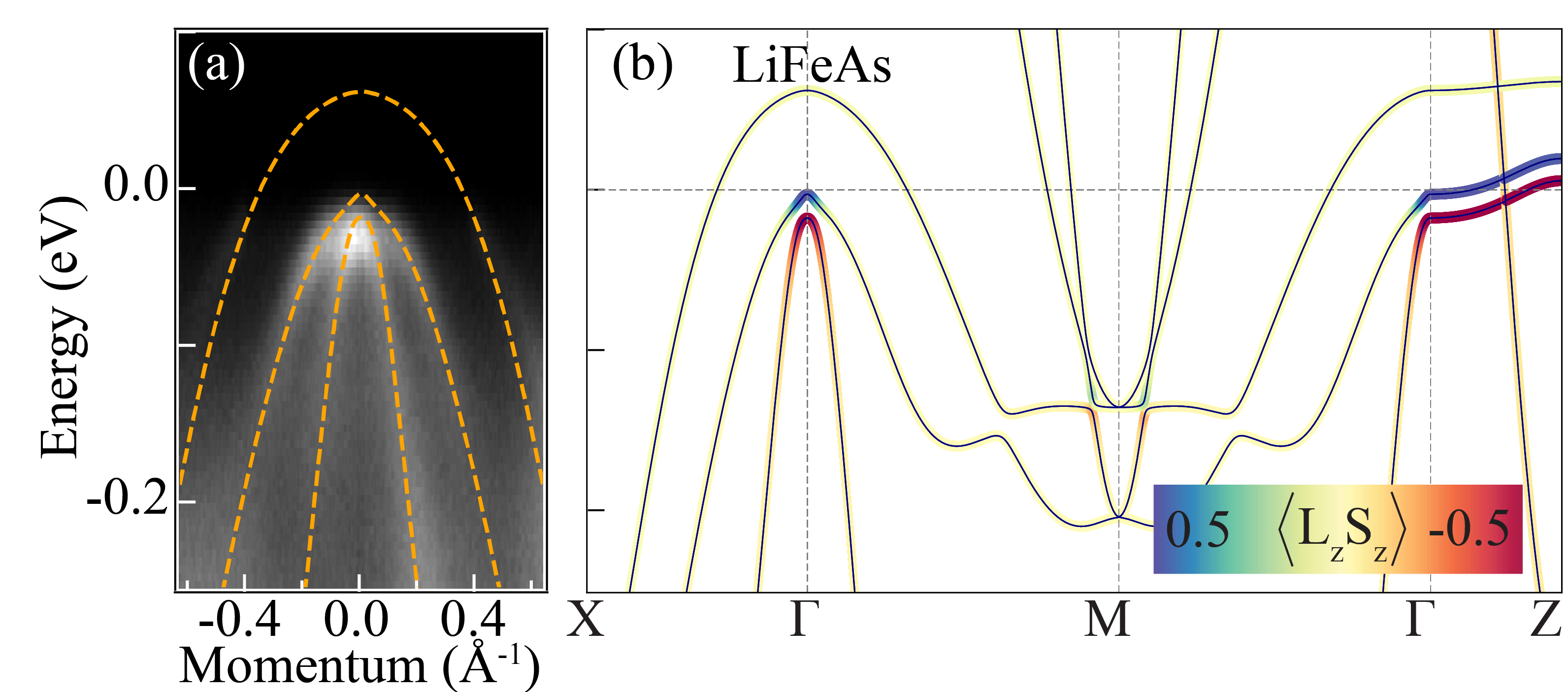}
\caption{(a) ARPES near normal emission for LiFeAs at $h\nu=$ 26 eV, corresponding to 0.1$\overline{\Gamma Z}$ with the TB model overlain in orange. (b) Tight-binding model for LiFeAs along high symmetry direction. Colourscale indicates the expectation value of $\left<L_zS_z\right>$.}
\label{fig:figLS}
\end{figure}

	In connection to the experiment, we plot the evolution of SOC in LiFeAs along high symmetry directions in Fig. 2 (b),2(c), emphasizing $\left<L_zS_z\right>$ due to its association with the measured $P_z$. To achieve agreement between the TB model and the ARPES dispersion, atomic SOC of strength $\lambda_{SOC}=18$ meV has been added to the Hamiltonian. The band dispersion observed in ARPES is renormalized by a factor of $\sim$2.2 from that of DFT, and so this SOC strength should be multiplied by the same factor to compare with DFT calculations \cite{Hirsch,vanr,renorm}. 

\begin{figure*}[ht]
\includegraphics[width=\linewidth]{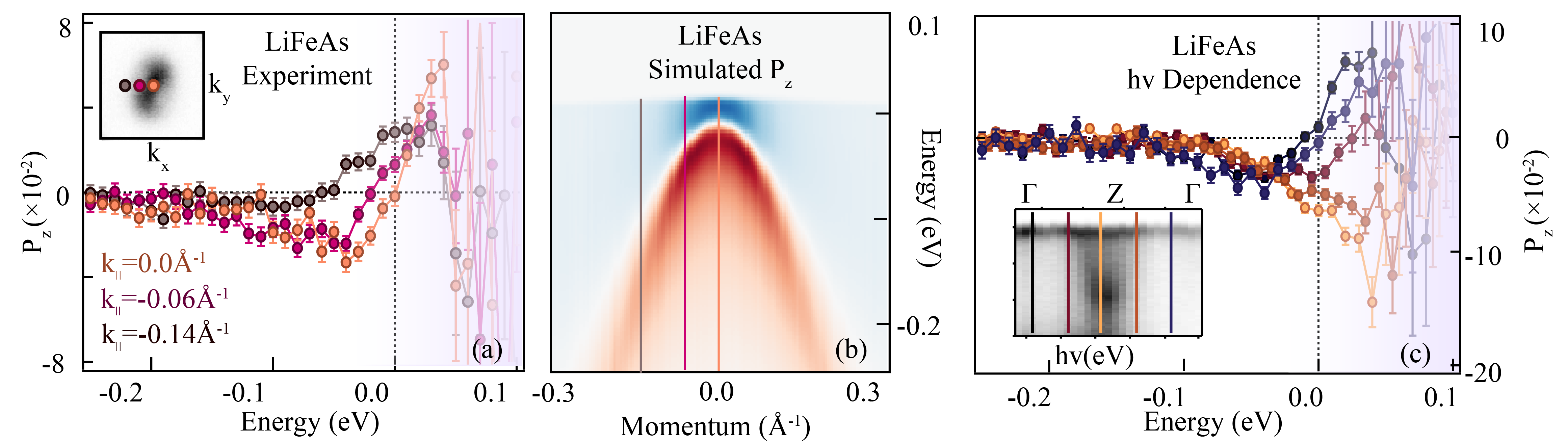}
\caption{(a) Measurement of out of plane Spin Polarization Asymmetry at normal emission (orange), $k_{||}$ = -0.06$\AA^{-1}$ (red) and $k_{||} = $ -0.14 $\AA^{-1}$ (purple). Inset shows the Fermi surface for this region of $\vec{k}$, with symbols indicating the momenta for the three curves. (b) The calculated map of $P_z$ (red minimum, blue maximum) near normal emission, vertical lines correspond to curves in (a). (c) Spin Polarization Asymmetry along $\overline{\Gamma Z}$. Spin polarization asymmetry was measured at $h\nu = $ 26 (black), 31(red), 36(yellow), 41 (orange), and 46 (purple) eV, corresponding to $k_z = 0.1Z$, $0.5Z$, $0.9Z$, $0.7Z$, and $0.35Z$ respectively. Inset: ARPES at normal emission as a function of photon energy--vertical lines indicate the photon energies (and $k_z$ values) for the spin measurements. }
\label{fig:figLiFeAs}
\end{figure*}
	CPS-ARPES was performed on LiFeAs, and for each emission angle we computed $P_z$ as in Equation \ref{eq:Peq}. The result is plotted in Fig. 3(a), where a switch in the sign of $P_z$ near $E_F$ is resolved, reflective of the switch in sign of $\left<L_zS_z\right>$ between the two doublets of opposite $\left<L_zS_z\right>$ from Fig. 2(b). This observation establishes for the first time an explicit correspondence between the splittings observed in standard ARPES experiments with spin-orbit coupling in these materials. Moving to larger $k_{||}$, the switch in $P_z$ moves with the dispersion to higher binding energies, and at large $k_{||}$, the amplitude of $P_z$ is also markedly reduced, consistent with our prediction of a concentration of the SOC effects near $k_{||}=0$.
	
	Simulated CPS-ARPES intensity across the entire region of momenta and energy near the zone centre allows for interpolation of $P_z$ throughout this volume of the Brillouin zone. In order to do this, simulated ARPES spectra were generated based on the experimental configuration (See Supplementary). The full simulated $P_z$ spectra is plotted in Fig. 3(b). As resolution and spin-incoherent background broaden and reduce the amplitude of the experimental $P_z$ (Fig. S3, S4), calculations facilitate comparison with $\left<L_zS_z\right>$. We infer that spin and orbital degrees of freedom in LiFeAs are coupled primarily near the zone centre where these bands approach the Fermi level. 
	
	In addition to spin-projection out of plane, we measured spin along the $\overline{\Gamma X}$ direction and found a small negative in-plane $P_x$ (Fig. S5), consistent with the $\left< L_xS_x\right>$ in Fig. S5(c). As suggested in context of Fig. 1, this requires hybridization with orbitals beyond $d_{xz}$ and $d_{yz}$, as SOC only introduces $L_zS_z$ terms between these states. This result should emphasize the importance of the full $\vec{L}\cdot \vec{S}$ operator in theoretical studies, as the $L_zS_z$ operator alone is insufficient to capture the nature of these states. Similar measurements (both in and out of plane) on the electron pockets at M produced no observable spin-orbital polarization (Fig. S6), suggesting the effects of SOC on the independence of $\vec{L}$ and $\vec{S}$ to be more relevant to the hole pockets at the zone centre. This has important implications for spin and orbital fluctuation pairing mechanisms which involve intra- and inter-band exchange in these channels \cite{Hirsch_pairing}, which are evidently co-dependent in regions of $k$-space relevant to superconductivity. Specifically SOC has been shown to suppress spin-susceptibility for $\vec{Q}$ connecting hole and electron pockets in FeSCs, of particular relevance to the viability of spin-fluctuation mediated superconductivity \cite{Kontani}.
	
	By tuning photon energy $h\nu$, we performed similar measurements along the third dimension ($k_z$) of the Brillouin zone \cite{Andrea}. As anticipated from the dispersion in Fig. 1(a) and 2(b), the outer hole-pocket moves well above $E_F$ towards $Z$. The spectrum is then dominated by the inner band with orbital and spin angular momentum aligned antiparallel, resulting in a strictly negative $P_z$ curve (Fig. 3(d)). By varying photon energy between 26 $eV$ and 46 $eV$, we followed the evolution of $P_z$ from $\Gamma$ to $Z$ and on to the next $\Gamma$.  The observed $P_z$ not only completes the momentum-dependence of SOC effects on the hole-bands, but also illustrates the sensitivity to the chemical potential: as the upper hole pocket moves above $E_F$, the corresponding Fermi surface is increasingly free of relativistic effects. In hole-doped FeSCs where hole band maxima are entirely above $E_F$, SOC effects will be suppressed. Similarly, in extremely electron-doped materials such as monolayer FeSe \cite{Ding}, the hole pockets are pushed below $E_F$, and the Fermi surface is defined in terms of non-relativistic electron pockets alone. 
	
	  In contrast to LiFeAs, many of the FeSCs undergo a nematic transition as temperature is reduced towards $T_c$. This has motivated theoretical interest in the role of nematicity in Fe-based superconductivity \cite{Alien}. In the nematic phase, we may ask if SOC is still of importance, or if the system is dominated by the energy scales associated with orbital ordering. In FeSe, there is a well-known structural distortion from tetragonal to orthorhombic around 90 K associated with the onset of orbital-ordering \cite{FeSe_growth}, providing the opportunity to explore SOC in the presence of nematicity. From ARPES measurements above $T_{ortho}\sim90$ K, the outer-hole band forms a small Fermi-surface, whereas the inner band has its maxima at 30 meV below $E_F$ (Fig. 4). CPS-ARPES confirms the origin of this splitting to be spin-orbit coupling. The energy scale of SOC is substantively larger than in LiFeAs, suggestive of perhaps increased hybridization with Se over As in the chalcogenide. It is instructive to note that while CPS-ARPES sacrifices resolution in contrast to modern ARPES (compare for example EDCs in Fig. 4(b) and 4(d), this is done to achieve an exceptional sensitivity to the spin-orbit induced polarization asymmetry, as exemplified by the sharp $P_z$ curve in Fig. 4(b). Below $T_{ortho}$, the dispersion of the two bands separates by an additional 10-15 meV along a range of momentum beyond $k_F$, demonstrating that SOC represents a larger energy scale than orbital order in this region of the Brillouin zone. At low temperature, we repeated momentum-dependent CPS-ARPES measurements as in LiFeAs, observing a similar evolution of the polarization asymmetry near zone centre as in Fig 3(a) (see Supplementary Fig. S7(a)). Near the Brillouin zone corner however, ARPES reveals more pronounced effects of orbital order\cite{FeSe_nem,Fanfarillo,Zhang_FeSe}, while CPS-ARPES recovers no substantive signatures of SOC(Fig. S7 (c)), illustrating the momentum-dependent interplay of these interactions.
	  
	  Through study of FeSe, we have demonstrated clearly that the orbital and spin degrees of freedom are coupled by relativistic effects more strongly than in the pnictides, and in a way which is not suppressed by the introduction of orbital order. This helps to justify the unanticipated INS results below $T_{ortho}$ in FeSe \cite{MaFeSe}, as the CPS-ARPES results demonstrate that SOC remains relevant in the presence of orbital order when $d_{xz}$ and $d_{yz}$ states are no longer degenerate in the absence of SOC. More generally, we have shown here via CPS-ARPES that albeit modest, spin-orbit coupling in FeSCs can result in a substantive modification of the electronic states and dispersion near the Fermi-level, and is therefore relevant to superconductivity and other low-temperature phases. While many interactions influence the low energy electronic dispersion, we have differentiated the influence of SOC from other perturbations such as orbital ordering. In the context of unconventional superconductivity, this further distinguishes the FeSCs from the strongly correlated cuprates, and rather is suggestive of comparison with the relativistic superconducting ruthenates \cite{Veenstra}. The FeSCs however, represent the possibility for supporting high temperature superconductivity in the presence of both correlations and relativistic effects. 

\begin{figure}[ht]
\centering

\includegraphics[width=\linewidth]{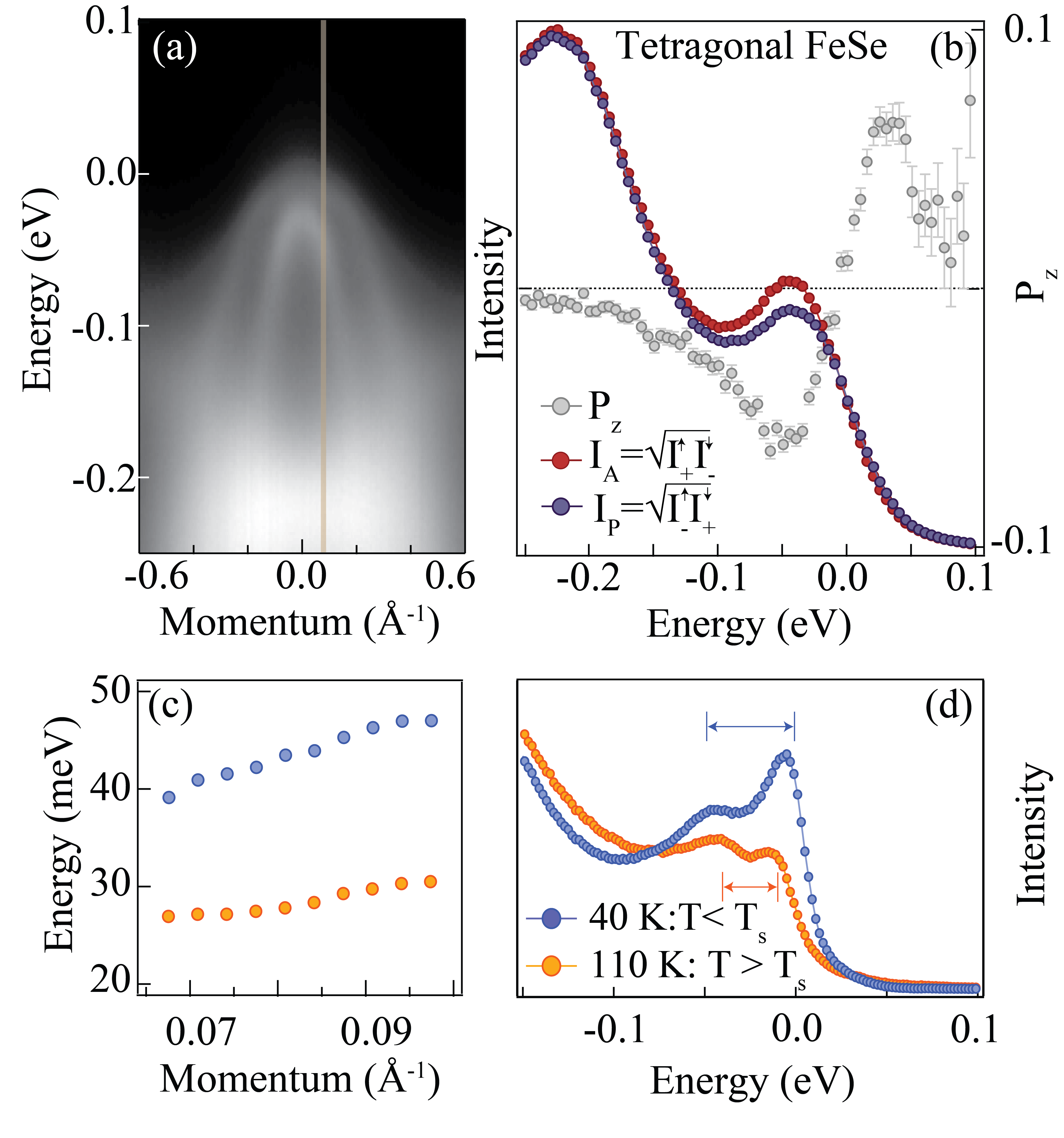}
\caption{a) ARPES image of FeSe at T=130 K (Sum of linear vertical and horizontal polarization maps at $h\nu=37$ eV). (b) VLEED Energy Distribution Curves (EDCs) for $I_P$ (purple) and $I_A$ (red), as defined above, measured near $k=0\AA$. The grey curve is the computed spin polarization asymmetry ($P_z$ of Equation 1) for these EDCs. (c) Energy splitting between hole bands for shaded region in (a) above (orange) and below (blue) the orthorhombic transition. (d) EDCs at $k_F$ above (orange) and below (blue) the orthorhombic transition temperature. Arrows illustrate the spread in peak positions across the transition. }
\label{fig:figFeSe}
\end{figure}
	
	Ultimately, the effect of SOC on FeSC phenomenology is not single valued and varies between materials. Our demonstration of the strong momentum-dependence of $\left<L\cdot S\right>$, and the consequent restoration of a non-relativistic description away from the Brillouin zone centre ($\Gamma Z$), emphasize the likely material-dependent influence of SOC. This point is manifest in, for example, the qualitative diversity of INS results \cite{MaFeSe}. Interestingly, the hole pockets where SOC effects are strongest are also associated in general with the largest reported superconducting gaps \cite{Borisenko}. As SOC has been suggested to suppress spin-fluctuation based pairing \cite{Kontani}, the sensitivity to the location of the chemical potential and the $d_{xy}$ band shown here suggests doping may act to mitigate the influence of relativistic effects, stabilizing spin-fluctuation mediated pairing. The situation is however not so straightforward as to label SOC as deleterious to superconductivity: one further consequence of SOC is the need to incorporate both spin-singlet and -triplet terms in the pairing equations \cite{Vafekgroup}, which in certain cases is necessary to stabilize attractive pairing in the s-wave channel \cite{VafekChub}. Despite the varied influence of SOC on phenomenology, our results present a common origin of relativistic effects in terms of the electronic structure; this will need be considered in attempts to further understand and manipulate the properties of FeSCs. 
	
	In conclusion, whether SOC is of more fundamental importance to the superconducting pairing mechanism in all FeSCs or rather has a more material-specific effect, remains to be determined. While discussions to date have primarily disregarded triplet terms and emphasized either orbital or spin-based fluctuation mechanisms supporting some type of s-wave superconductivity, the results here suggest that further consideration of the pairing mechanisms put forth thus far and their possible interplay will be needed for a more complete understanding of superconductivity in the Fe-pnictides and chalcogenides. 

We thank O. Vafek for helpful discussions on the topic. This research was undertaken thanks in part to funding from the Max Planck-UBC-UTokyo Centre for Quantum Materials and the Canada First Research Excellence Fund, Quantum Materials and Future Technologies Program. The work at UBC was supported by the Killam, Alfred P. Sloan, and Natural Sciences and Engineering Research Council of Canada (NSERC) Steacie Memorial Fellowships (A.D.), the Alexander von Humboldt Fellowship (A.D.), the Canada Research Chairs Program (A.D.), NSERC, Canada Foundation for Innovation (CFI) and CIFAR Quantum Materials Program. E.R. acknowledges support from the Swiss National Science Foundation (SNSF) grant no. P300P2-164649. This work has been partly performed in the framework of the nanoscience foundry and fine analysis (NFFA-MIUR Italy Progetti Internazionali) facility. 

\bibliography{SOC_FeSC}

\end{document}